\documentclass[12pt]{article}

\usepackage{amsmath,amsfonts,amsbsy,amssymb,fullpage,makeidx,array,graphicx,epsfig,enumerate}

\makeindex
\newcounter{defncntr}

\newcounter{thmcntr}

\newcounter{remcntr}

\renewcommand{\theremcntr}{\arabic{section}.\arabic{remcntr}}

\newcounter{lemcntr}

\newcounter{egcntr}

\begin{document}

\thispagestyle{empty}

\title{Entanglement and Thermodynamics \\ of Black Hole Entropy}

\author{ R. Brout\thanks{Email: robert.brout@ulb.ac.be}\\ \\[8mm]
\small Perimeter Institute for Theoretical Physics\\ \small Waterloo,
Ontario N2L 2Y5, Canada\\[8mm] \small
Departments of Applied Mathematics and Physics \\ \small University of
Waterloo, Waterloo,
Ontario N2L 3G1, Canada \\[8mm] \small
Service de Physique Th\'eorique, Universit\'e Libre de Bruxelles \\
\small The International Solvay Institutes\\
\small B1050 Bruxelles, Belgium}

\date{}

\maketitle

$$$$$$$$
\begin{abstract}

Using simple conditions drawn from the stability of the cosmos in terms of
vacuum energy density, the cut-off momentum of entanglement is related to
the planckian mass.  In so doing the black hole entropy is shown to be
independent of the number of field species that contribute to vacuum
fluctuations.  And this is in spite of the fact that the number of field
species is a linear multiplicand of the entanglement entropy when this
latter is expressed in terms of the fundamental momentum cut-off of all
fields.
\end{abstract}

\newpage

An interesting problem of fundamental character arises upon comparing
black hole entropy, $S$, as determined thermodynamically by Hawking [1]
from black hole evaporation, and a statistical procedure called
entanglement entropy [2, 3].  The two ought to be the same if they do
indeed describe the same physics.  The formulae to be compared are:
\begin{align}
S &= m^{2}_{~pl} A \qquad \text{(Hawking)} \tag{1a} \label{e1a}\\
&= \nu \Lambda^{2} A \qquad \text{(Entanglement)} \tag{1b} \label{e1b}
\end{align}
All constants of $O(1)$ are set equal to $1$. $A$ is the area of the black
hole horizon, $m_{pl} =$ planck mass, $\Lambda$ is a momentum cut-off
which has been introduced to implement the counting procedure that leads
to equation (1b). Also, $\nu$ is the number of species of fields which is
introduced to describe the field vacuum fluctuations.

One's first inclination would have been to set $\Lambda = m_{pl}$ since
$m^{-1}_{~pl}$ is the fundamental length scale of gravitation; and indeed
of all physics.  Since it is thought to describe geometry, $m^{2}_{~pl} R
\sqrt{g}$ being the gravitational action without reference to matter,
$m_{pl}$ should contain no reference to $\nu$.  Therefore, consistency
requires that we give up our first inclination and attribute a dynamical
character to $\Lambda$ wherein \setcounter{equation}{1}
\begin{equation}
\Lambda^{2} = m^{2}_{~pl}/\nu. \label{e2}
\end{equation}
This short essay contains a simple argument which we believe has the germs
of a rigorous derivation of equation \eqref{e2}.  It is to be noted that
equation \eqref{e2} has far reaching implications in our appreciation of
quantum field theory.  $\Lambda$ is not a momentum cut-off which has been
conveniently introduced as a regulator to make calculations possible.
Rather it is a dynamical parameter precisely determined through use of the
laws of physics.  We shall show that it comes about from the existence of
a stable cosmos.

Let us begin by reconsidering the origins of equations (1a),(1b).  First,
equation \eqref{e1a} follows from the existence of a black hole
temperature, a concept that arises from the periodicity of asymptotic
field Green's functions in imaginary time.  The temperature is the inverse
of that period.  Dimensional arguments suffice to deduce $T_{BH} =
m^{2}_{~pl}/M$ ($T_{BH}=$ black hole temperature; $M =$ the black hole
mass).  The radius of the black hole is $M/m^{2}_{~pl}$, hence $A =
M^{2}/m^{4}_{~pl}$.  Integration of $dS = dE/T = dM/T$ then gives equation
\eqref{e1a}.

Equation \eqref{e1b} is conveniently derived from the partially traced
density matrix that describes field configurations about the black hole.
Since $S$ is asymptotic in character (i.e., far from the black hole) it
suffices to consider field modes in flat space.  Therefore one can
idealize.  Consider a cubic lattice of cells $\Lambda^{-3}$ in dimension.
Divide the cube into two parts, large and small, by a plane at, say $Z=$
constant.  The field can be modelled as a set of springs coupling
neighbouring points, so we have left $(L)$ and right $(R)$ field
configurations.

To seize the meaning of entanglement entropy first imagine removing all
the springs directed along the $Z$ axis that are bisected by the dividing
plane.  Then, $L$ and $R$ are decoupled and the Schr\"{o}dinger
representation of the ground state factorizes into two ground states,
$\Psi = \Psi_{L} \Psi_{R}$ where, $\Psi_{L} (\Psi_{R})$ refer to degrees
of freedom in the $L(R)$ sectors.  Then $S = S_{L} + S_{R} = 0 + 0$.

A non trivial entropy  can be constructed by reinstating the coupling of
$L$ and $R$ by replacing the missing springs and forming a reduced density
matrix, $\rho_R$, by tracing over the $L$ degrees of freedom, i.e.,
$\rho_{R}=tr_{L}\rho$, where $tr_{L}$ is the trace over $L$ degrees of
freedom. The partial trace is carried out to express one's ignorance of
the field configurations within the black hole.  Because of the existence
of the springs at the $LR$ boundary, the entropy $S_{R}$ no longer
vanishes, where $S_{R} = - tr_{R} \rho_R \ln \rho_{R}$.  It is called
entanglement entropy, the $L$ degrees of freedom being inevitably tangled
with those on the right because of the ``bridge'' springs across the
boundary.

The number of such bridge springs per particle species is $\Lambda^{2}A$
so the induced effect is expected to be $S_{R} = \nu \Lambda^{2} A$.  The
factor $\nu$ arises from the $\nu$ species appearing in the trace.  The
argument of proportionality to $A$ is correct, as such, only if there are
no long range correlations.  This is true if the particle at each site has
a mass.  A rigorous calculation [2,3,4] however shows that a mass is, in
fact, unnecessary to complete the calculation and equation \eqref{e1b} is,
in general, the correct answer.  As the details of this proof are not
germane to our present purpose we refer the reader to the references.

The point of this note is to show how simple, yet fundamental, reasoning
leads to the consistency condition \eqref{e2}.  It is based essentially on
the stability of the cosmos as deduced from quantum field theory (QFT) and
general relativity (GR) applied to homogeneous flat spaces.  These are the
spaces that one generally considers asymptotically far from the black
hole.

We begin with a highly oversimplified estimate for the vacuum energy
density of such spaces for bosonic fields, based on the classification of
field configurations in terms of modes.  There are $\nu$ elementary
fields, each developed in terms of modes and all cut-off at a common value
of $\Lambda$ (once more all factors $O(1)$ are set $=1$).
\begin{equation}
\rho = \nu \Lambda^{4} - \frac{\nu^{2} \,\Lambda^{6}}{m^{2}_{~pl}} \label{e3}
\end{equation}
The first term on the right hand side is the zero point energy calculated
to the lowest non-vanishing order in $m^{-2}_{~pl}$ (i.e. independent of
$m_{pl}$).  In addition there is the universal interaction among all
fields, mediated by gravity. Since this effective interaction is
attractive, the corresponding energy is negative. A crude approximation is
a pairwise interaction, such as the Newtonian potential, as expressed by
the second term on the RHS of Eq.3, to order $O(m^{-2}_{~pl})$.  Note that
all species interact universally through gravity, hence the combinatorial
factor of $\nu^{2}$, with the minus sign expressing the attractiveness of
gravity.  The dependence on $\Lambda$ follows from dimensional arguments.

In the adiabatic era one has $\rho_{\text{Total}}=\rho_0+\rho_M$ where
$\rho_M$ is the energy-density due to on-mass-shell quanta and $\rho_0$ is
the vacuum energy, associated by most physicists, with dark energy. Whether $\rho_0$ is
strictly positive or a quantity that fluctuates about zero mean, it cannot,
in absolute value, exceed $H^2$ in order of magnitude. At the present time
this is $O(10^{-100}m^4_{pl})$ whereas each of the two contributions to
the r.h.s. of Eq.3 are $O(m^4_{pl}/\nu)$. Therefore, in the adiabatic era,
in good approximation the separate terms contributing to $\rho_0$, given
by Eq.3, cancel. And Eq.2 is secured. If a black hole during inflation the
situation could be more complicated and will not be discussed here.

Before trying trying to evaluate the validity of the estimate given by
Eq.3, it is meet that the reader appreciate the deep cosmological
significance inherent in equation \eqref{e3}.  To this end, it is
convenient to envision $N(\Lambda) = \text{const.} \Lambda^{3}$, the
density of modes, as an analog to the particle density of a quantum fluid
which is self interacting through an attractive pairwise interaction.
Whereas in conventional quantum fluids the ground state and its various
concomitant dynamical parameters are determined variationally, this is not
true of the ``cosmic fluid''. Rather, the usual appeal to Pauli repulsion,
which prevents total collapse, is replaced by the positivity condition,
$\rho \geq 0$.  As we have pointed out, in the adiabatic era this is
tantamount to $\rho = 0$, thereby leading to equation \eqref{e2}.  In a
longer follow-up paper, these considerations will be extended to lead to
further understanding of mode dilution, fluctuations such as dark energy,
inflation and its fluctuations, and other features of cosmology. For the
nonce, we merely wish to convey the message of equation \eqref{e2}, a
highly nontrivial condition based on cosmic stability.
\medskip\newline
\noindent Let us now delve somewhat into the nature of the approximations inherent
in Eqs.1a,1b,3.
\smallskip\newline
Equation 1a) is in the nature of a thermodynamic identity given the
classical black hole metric, i.e., the neglect of backreaction occasioned
by evaporation as well as the possible effect of fluctuations of the
horizon. That the temperature is unaffected by self-interaction of the
field is a well-known theorem. However, this has not been checked in the
case that the interactions are gravitational. This essay is not concerned
with these questions and Eq.1a is accepted as such.

Equation 1b) has been calculated using free field theory [2,3]. Hence its
statute is different from equation 1a). To compare equations 1a) and 1b)
is, therefore, analogous to the fleshing out of a thermodynamic identity
with a formula derived from a kinetic model, here free field theory.
Equation 2) is thus to be regarded as a consistency condition subject to
the qualification that the free field theory is applicable to
entanglement.

Accordingly, one should not push Eq.3 too hard. Each of the two terms is
estimated to lowest order in $m^{-2}_{pl}$ and one expects there to be
corrections. It is also to be mentioned that the identification of
$\Lambda$ with a sharp momentum cutoff is an over-idealization. Rather,
one expects $\Lambda$ to be ``fuzzy". This is because the considerations
leading to Eq.3 show that $\Lambda$ is determined from an equilibrium
between zero point energy of modes and their gravitational interaction
energy. This equilibrium fluctuates, hence the ``fuzzyness". On this
basis, exact agreement of Eq.1a and Eq.1b to terms of $O(1)$ must be
imposed on grounds of thermodynamic consistency. Therefore, this essay is
to be considered only as a semi-quantitative explanation of Equation 2.

If $\nu\gg 1$, an expansion in $\nu^{-1}$ can be carried out. This must be
done for Eq.1b as well as each of the terms contributing to Eq.2. Of
these, the interaction term of Eq.2 is immediately evaluated to leading
order in $\nu^{-1}$, being a sum on simple loops. Each loop carries a
factor $\nu m^{-2}_{Pl}\Lambda^2$. This, being $O(1)$ does not affect the
previous estimate at the precision given. It may be conjectured that the
same is true for the other terms, but this remains to be carried out.
\smallskip\newline
We close this essay with some relevant references. Parentani [5], in a
series of works has shown that modes of sufficiently short wavelength are
over-damped due to their scattering from vacuum fluctuations, thereby
corroborating the idea that $\Lambda$ is dynamically generated.

Effects of renormalization are extensively discussed by Jacobson and
collaborators. See [6] and further references therein.

Susskind and Uglum [7] have investigated black hole entropy through use of
the deficit angle formalism.

It will be interesting to interrelate these various approaches to the
simple physical picture put forth in this essay.
$$$$
\bf Acknowledgement. \rm Grateful acknowledgement is extended to William
Donnelly, Ted Jacobson, Achim Kempf, Simone Speziale and Larry Susskind
for helpful discussions.

\subsubsection*{Bibliography:}

\begin{itemize}

\item[{[1]}]
S.W. Hawking, Commun. Math. Phys. 43, 199 (1975).

\item[{[2]}]
L. Bombelli, R.K. Koul, J. Lee, R. Sorkin, Phys. Rev. D, 34, 373 (1986).

\item[{[3]}]
M. Srednicki, Phys. Rev. Lett., 71, 666 (1993).

\item[{[4]}]
F. Piazza, F. Costa, arXiv:0711.3048

\item[{[5]}] R. Parentani, arXiv: 0710.4664, 2007

\item[{[6]}] C. Eling, R. Guedens, T. Jacobson, Phys. Rev. Lett., 96, 121301 (2006)

\item[{[7]}] L. Susskind, J. Uglum, Phys. Rev. D 50, 2700 (1994),
hep-th/9401070

\end{itemize}
  \end{document}